# Skill Downgrading Among Refugees and Economic Immigrants in Germany: Evidence from the Syrian Refugee Crisis[☆]

Plamen Nikolov[a,d,e]    Leila Salarpour G.[a,b]    David Titus[c]

This version: October 18, 2021

**Abstract:** Upon arrival to a new country, many immigrants face job downgrading, a phenomenon describing workers being in jobs far below where they would be assigned based on their skills. Downgrading leads to immigrants receiving lower returns to the same skills than natives. The level of downgrading could depend on the type of immigrant and numerous factors. This study examines the determinants of skill downgrading among two types of immigrants – refugees and economic immigrants – in the German labor markets between 1984 and 2018. We find that refugees downgrade more than economic immigrants, and this discrepancy between the two groups persists over time. We show that language skill improvements exert a strong influence on subsequent labor market outcomes of both groups. (*JEL* J11, J15, J61, F22, O15, J61)

*Keywords*: downgrading, immigrants, refugees, Germany, labor markets, wages, employment

[☆]We are grateful to the research staff at Deutsches Institut für Wirtschaftsforschung (DIW), Berlin who made the SOEP data available to us and provided numerous insights based on their field experience implementing the survey. We thank Matthew Bonci, Martin Farnham, Sol Polachek, and Andreas Pape, for constructive feedback and helpful comments. All remaining errors are our own.

[☆]Contact information: Titus: Department of Economics, Cornell University. E-mail: dwt45@cornell.edu; Salarpour: Department of Economics, University of Victoria. E-mail: lsalarpour@uvic.ca. Nikolov: Department of Economics, State University of New York (Binghamton). E-mail: pnikolov@post.harvard.edu.

[a] State University of New York (at Binghamton)
[b] University of Victoria
[c] Cornell University
[d] IZA Institute of Labor Economics
[e] Harvard Institute for Quantitative Social Science

# 1. Introduction

Immigration into Germany skyrocketed in the 2010s (BPD, 2018) due to liberalized migration policies and the 2011 Syrian refugee crisis. The economic performance of immigrants has significant economic and social consequences for the host country and is at the heart of numerous immigration debates. Among the issues of particular importance is the so-called skill downgrading, which occurs when immigrants obtain lower economic returns to their skills than comparable native workers. Downgrading implies that immigrants receive lower wages for the same skills than natives. Immigrants could also skill downgrade in a different sense: they work in different occupations than natives even if they hold comparable education and work experience.

Extensive empirical literature documents the magnitude of the downgrading phenomenon in various high-income settings (Eckstein and Weiss, 2004; Dustmann, Frattini, and Preston, 2013; Dustmann and Preston, 2012; Kossoudji and Cobb-Clark, 2000; Borjas, 2003). A better understanding of the magnitude of skill downgrading is essential because it can generate considerable labor market frictions in the employment outcomes of immigrants – job-skill mismatches. Moreover, the presence of downgrading can significantly influence how immigration inflows affect lower-skill native workers. Numerous studies aim to assess the immigration impacts on labor market outcomes by classifying workers into national skill-cell or education-experience categories (Breell et al., 2020).[1] If immigrants downgrade, obtaining estimates based on the skill-cell approach and the mixture approach on how immigrants affect the labor market outcomes of natives will be biased.

---

[1] Studies classifying immigrant inflows into education-experience cells on a national level rely on the so-called "national skill-cell approach." An alternative econometric approach uses variation in the total immigrant flow into a country based on regions ("spatial approach"). A third approach relies on variation in immigrant inflows across education groups and regions (a "mixture approach").

This study examines the economic performance of two types of immigrants in Germany: economic immigrants and refugees. An important characteristic distinguishing the two groups is that economic immigrants typically opt to migrate if their migration maximizes household income and could return to their country of origin. In contrast, refugee immigrants migrate primarily for political reasons. Given the distinct characteristics of the two groups, we specifically examine for differences in the magnitude of skill downgrading. Significant downgrading for either group has crucial economic implications for the immigrants' welfare and has potential implications for productivity losses in the host country if significant job-skill mismatches occur. First, we focus on the degree of downgrading differs between these two groups. Second, we attempt to shed light on how the two groups differ in the assimilation process. We specifically focus on determining the degree of skill downgrading and investigating whether wage convergence differs between the two groups.

We conduct our empirical analysis using the German Socio-Economic Panel (SOEP), an annual panel survey of households and individuals. The SOEP collects rich longitudinal data on labor market outcomes for natives and various immigrant groups in Germany. Our analysis uses the SOEP main and migration samples to study labor market outcomes of immigrants who arrived in Germany between 1995 to 2018. Our empirical strategy investigates how wages and employment outcomes vary for different immigrant groups, accounting for differences in human capital, work experience, occupational features, and other labor market characteristics.

Our analysis reveals several findings. First, consistent with the existing studies, we show that both economic and refugees downgrade significantly upon arrival. However, we show that refugees downgrade differently based on gender: females downgrade more in hourly wages (conditional on working) but less in employment. Second, despite the initial downgrading, we find evidence of refugees converging to the other immigrant groups regarding wages. Male refugees do not converge at all in hourly wages, while female refugees converge relatively quickly. Third, we shed light on the significant role that language acquisition can exert in boosting worker wages of immigrant groups. With respect to employment status changes, we show that improving language abilities carries a higher return for refugees than economic immigrants among the male sample. However, female refugees have no extra return in employment compared with female economic immigrants. Finally, regardless of gender, our analysis reveals that refugees do not receive extra wage returns to language skill improvements compared to economic immigrants.

Our study advances the literature in two significant dimensions. In contrast to previous studies (Cortes, 2004; Ruiz and Vargas-Silva, 2018), we show that gender the magnitude of skill downgrading could hinge critically on gender. Second, we shed light on the role of various factors that help particular immigrant groups converge in wages and employment status relative to economic immigrants and natives.

The remainder of the paper proceeds as follows. Section II overviews the related literature and unresolved issues. Section III describes the data. Section IV outlines the empirical strategy. Section V presents the results. Section VI concludes.

## 2. Literature Review

### A. Causes of Migrant Choices for Economic Immigrants and Refugees

A core feature of economic migration models relates to what motivates people to migrate. The Roy model, for instance, predicts that immigrants self-select into nations where they expect to earn more than they would in their home country.[2] However, this framework may not apply to all immigrant groups. Refugees do not migrate for economic reasons. Instead, they migrate as a consequence of violence or oppression in their country of origin. Since their top priority is to find a safe country that will accept them, refugees may not reach a destination that fits their prior education and work experience or produce an optimal job-skill match. Moreover, refugees are unlikely to prepare to locate an optimal job-skill match in the new labor market. Borjas (1982) contends that these factors imply that refugees have lower initial returns to their human capital than economic immigrants with similar characteristics.[3]

---

[2] See Borjas (1987) for a detailed application of the Roy model to immigration.

The differences between economic immigrants and refugees carry over to language acquisition. Since refugees do not prepare for life in a new country before migrating, they may lag behind other immigrants' initial host country fluency. Although Cortes (2004) observes no differences in host country fluency between immigrants and refugees in the US, Brell et al. (2020) find that refugees in Europe begin with worse language skills than other new immigrant groups. Thus, refugees in countries like Germany may face higher language barriers than their economic counterparts. Although the exact impact of language is difficult to estimate, studies consistently find that poor language fluency has adverse impacts on immigrants' labor market outcomes.[4] Chiswick and Miller (2002) report that immigrants with low English fluency in the U.S. have much lower returns to education. Cortes (2004), Yao and Ours (2015), and Lessem and Sanders (2020) find that better fluency increases earnings. Meanwhile, Kostenko et al. (2012) assert that language ability is crucial in preventing occupational downgrading. Thus, refugees may be set further back in the labor market through lower fluency than other immigrants.

B. Immigration Barriers for Economic Immigrants and Refugees

In addition to the language obstacle, occupational and employment regulations affect economic immigrants and refugees differently. Market barriers – occupational licensing and work permits – can generate additional market frictions.[5] Moreover, refugees face additional restrictions. Many countries in Europe, including Germany, temporarily bar refugees from employment upon arrival, influencing their short-run and long-run performance in the labor market.[6]

---

[3] Borjas (1982) explains this point further, likening refugees to workers who are unexpectedly laid off, compared to other workers who plan to quit and re-enter the labor market.

[4] As articulated by Dustmann and Van Soest (2001), measurement error can bias results, and correlation between language acquisition and unobserved ability can lead to endogeneity problems. It is challenging to design an exogenous instrument for language, but studies using instruments often reach different results from OLS. Chiswick and Miller (1995) find that OLS underestimates the impact of language on earnings. When Yao and Ours (2015) use OLS, they record only small penalties for poor language ability for both men and women in the Netherlands. But when they instrument for language with an interaction of age at arrival with languages spoken as a child, they discover a 48 percent earnings penalty for women. Dustmann and Fabbri (2003) do not find different results from OLS with their instrument when testing employment, but they do find divergent results when regressing on earnings.

Refugees and economic immigrants may also differ in their choice of location within the host country. Immigrants who seek to maximize their earnings tend to settle around "thicker" labor markets. Refugees, however, often have limited choices in where they initially settle. In Germany, for example, the government disperses refugees into districts quasi-randomly based on a rule, the Königsteiner Schlüssel formula. Because of this policy, refugees cannot self-select into more propitious districts, at least not in the short run. The placement rule makes it challenging for refugees to settle in co-ethnic areas. When migrants sort into areas exhibiting higher shares of their ethnic groups, this pattern of migration sorting can lead to beneficial outcomes. Beaman (2012) finds that immigrants have a higher employment probability in ethnically concentrated areas, primarily due to immigrants forming better social networks conducive to subsequent employment opportunities. Simmilarly, Damm (2014) finds that immigrants in Denmark have higher employment rates in areas with higher fractions of ethnic minorities as a share of the overall population.

Refugees do not prepare for their relocation and meet higher regulations and limited location choices in their host countries. Therefore, some studies show that refugees downgrade more than other immigrant groups. Cortes (2004) finds that both male and female refugees in the U.S., initially earn less than their non-refugee counterparts. Similarly, Ruiz and Vargas-Silva (2018) demonstrate significant earnings and employment gaps between economic immigrants and asylum seekers in the U.K. Unlike Cortes (2004), Ruiz and Vargas-Silva (2018) find that the gap is wider for females than males. Existing studies show a consistent finding: the two immigrant groups exhibit an initial gap in employment and earnings.[7]

---

[5] Lessem and Sanders (2020) find that eliminating occupational barriers in the US would increase the initial earnings of high-skilled immigrants by 38 percent.
[6] Brell et al. (2020) discuss the impact of temporary employment bans in more detail.

Existing studies demonstrate that refugees fare worse in the short-term labor market outcomes than other immigrant groups. However, what remains unclear is whether these initial disadvantages persist in the long run. Furthermore, no existing studies investigate what factors can help mitigate the disparity between the two immigrant groups. Although there is widespread agreement that refugees downgrade more than other immigrants upon arrival, there is no consensus on whether the two groups converge in employment and earnings over time. Cortes (2004) observes that refugees in the U.S. catch up to (and even surpass) economic immigrants in earnings after ten years. The study attributes this pattern to the expected length of stay.[8] Dustmann et al. (2017) find that refugees in the E.U. catch up to other immigrants in employment. In contrast to these two studies, Brell et al. (2020) discover that, while refugees converge towards immigrants, the employment and earnings gap persists in many European countries. Ruiz and Vargas-Silva (2018) report convergence in employment in the U.K. but a persistent gap in earnings. The long-term outlook for refugees likely varies by nation.[9]

Although existing studies have examined the initial gap between the two groups, the long-run trajectory and its underlying causes remain unclear. Our empirical analysis investigates whether refugees subsequently upgrade and converge in their earnings and employment. We also provide speculative evidence on some of the mediating channels behind these patterns.

---

[7] See also Dustmann et al. (2017), Neuman (2018), Brell et al. (2020).
[8] Assuming that it is more difficult for refugees to return to their home country, Cortes contends that they invest more in host country-specific human capital than other immigrants. This point is also made by Borjas (1982). Corroborating this theory is Cortes's finding that refugees in the US rise faster in educational attainment and become more fluent in English over time relative to their non-refugee counterparts. However, Dustmann et al. (2020) note that long-term refugees in Europe lag behind other long-term immigrants in fluency.
[9] Another explanation for these disparities is the use of panel data by some, compared to cross sectional data by others. Using cross sections over long periods allows new cohorts to enter the sample and bias the results. This problem is most notably articulated by Borjas (1985). Ruiz and Vargas-Silva (2018), who use cross sectional data, hypothesize that this may cause the disparity in their results with the panel-level approach by Cortes (2004).

## 3. Data and Summary Statistics

### A. Data Sources

**The German Socio-Economic Panel (SOEP).** Our analysis draws on data from the German Socio-Economic Panel (SOEP). The SOEP, an annual panel survey of households and individuals in Germany, follows a similar structure to the U.S. Panel Study of Income Dynamics. The SOEP was first conducted in 1984 in West Germany with about 4,500 households. After the German reunification, the survey added 2,170 households from East Germany in 1992. The survey supplemented the core sample with refreshment samples in 1998 and 2000, substantially increasing the total survey sample size. The current survey covers 30,000 individuals from 15,000 households. The panel follows individuals at the household level. However, all participants aged 16 years or older go through face-to-face interviews. We use data drawn from survey waves conducted between 1995 and 2018. However, we focus on individual labor market outcomes. Therefore, we analyze labor market outcomes for years before the survey (a particular survey year elicits employment data for the prior year).

An attractive feature of the SOEP is that the survey collects rich longitudinal data on labor market outcomes for immigrants in Germany. In 2013, the survey added an immigrant sample of 5,000 persons. The new immigrants were sampled based on the 2011 Census to account for compositional changes since 1995. In addition to standard demographic and socioeconomic characteristics, the survey also collects information on other variables we analyze: German-language skills, the country of birth, and current location in Germany.

**SOEP Data on Immigrants and Refugees**. In addition to the primary SOEP sample, we also draw on from the IAB-SOEP Migration Sample. This additional dataset contains data highly suitable to our main research question. The survey collects data, from 4,964 people in 2,723 households, on the living situations of new immigrants to Germany. The survey collects information on immigrants, including their country of origin, pre-immigration education, labor market status, and family background characteristics. In addition to various human capital proxies (i.e., educational attainment), the migration sample elicits German skills in speaking, reading, and writing at two points in time: upon their arrival in Germany and at the time of the survey. Each skill was self-assessed by the respondents using a Likert scale ranging from excellent (5), good (4), sufficient (3), poor (2), and none (1). Because the migration sample is a panel, the survey collects data on immigrants over time, allowing us to observe changes across different periods.

This additional sample also has several critical advantages suitable for our empirical analyses. First, the individual data is linked to register data from the Integrated Employment Biographies (IEB) database of the Institute for Employment Research (IAB), which contains rich labor market history of individuals in Germany. This linkage provides information on wages and salaries, employment, unemployment and benefit receipt, and many other variables that are particularly relevant to labor market issues.

In addition to labor market data, the migration module details the complete migration, education, and labor force participation histories in the country of origin and any other countries of residence. Such data can enable a thorough analysis of significant life events in the respondent's home country, Germany, or other countries. In addition, the data can enable economic analyses of the integration progress into the German labor market and other societal aspects. The migration questionnaire also includes several new questions on earnings and labor market integration and occupational status before migration, migration decisions in the family and partnership context, and purposes and transfer channels of remittances.

We only include employable persons aged 16–65. We restrict the sample to first-generation immigrants born abroad and exclude individuals with missing information about their year of migration, language skills, and residence status. There are six different categories of immigrants: family members, asylum seekers, ethnic

Germans, students and apprentices, immigrants who arrived as job searchers in Germany, and immigrants with a job commitment, whereas the two latter groups are at the center of attention in our empirical analysis. We code immigrant status as a binary indicator; we define immigrants based on information regarding a foreign birth. We also create a variable capturing year since immigration, calculated as the difference between the survey year and the year of arrival in Germany. This variable is a proxy for the experience gained in the country and opportunities for social integration.

In our analysis, we distinguish two categories: economic immigrants and refugees. We define an economic immigrant as someone not born in Germany but residing in Germany and not explicitly flagged as a refugee in the survey. The group of refugees comprises individuals whom the SOEP defines as either asylum seekers or refugees. Most of our refugee observations are newer arrivals, with a big jump in refugees surveyed in 2016. About 7 percent of immigrants in our sample, the vast majority of whom are non-refugees, have no reported year of migration and are excluded from our analysis.

**Labor Market and Other Socioeconomic Outcomes.** For our empirical analysis, we use data on employment status and earnings. We define employed individuals as those who report currently working. To minimize potential measurement error in employment classification, we classify individuals whose hourly wage rate falls below a particular benchmark (i.e., less than 3 euros per hour) as not employed.[10] We do not include self-employed head of households as the reported earnings of this group has a capital income component. We transform monthly gross earnings into hourly gross earnings using data on the working hours per week.

B. Descriptive Statistics

---

[10] This approach rules out some observations with implausibly low values for earnings.

First, we start with an overview of the analysis sample. Table 1 reports the number of observations aged 16-65 by type and years since migration. The table includes observations from immigrants interviewed in multiple years. In total, the sample comprises 13,606 economic immigrants. We analyze 90,941 observations from these immigrants. There are 9,304 unique refugees, with 19,837 observations. The SOEP initially surveyed eight thousand sixty-six refugees within their first five years in Germany, and most of these come from more recent refugee cohorts. Both refugees and economic immigrant observations have a similar distribution of males and females over time.

[Table 1 about here]

Table 2 reports the number of immigrants aged 16-65 by country of origin and immigration period.[11] The number of refugees has significantly increased in the analysis period: 8,007 of the 9,304 refugees entered Germany in 2011 or later. Unlike previous cohorts, the most recent refugee cohort hails from a Middle East country. Refugees from Syria in the most recent cohorts constitute a significant source of immigration inflows. Economic immigrants come from a broader set of countries. The majority of our economic immigrant sample entered Germany before 1995.

[Table 2 about here]

Table 3 details the summary statistics of our analysis sample.[12] We report data on native Germans for comparison purposes. As the numbers reveal, compared to the native population, immigrants have fewer years of schooling, report smaller family sizes, live in larger households, and are more likely to be married than natives.

---

[11] Although the legal working age in Germany is 15, our sample has no relavent data on anyone below age 16.
[12] Religious affiliation is not well-populated data. However, based on the available data, immigrants tend to be more Christian, while refugees are from Muslim backgrounds. This follows the pattern of immigration presented in table 2 and the origin country of immigrants.

In addition to differences between Germans and immigrants, the table reports descriptive statistics on economic immigrants and refugees. In terms of age, refugees are six years younger on average than economic immigrants. Refugees have a higher number of children living in the household. Moreover, unlike economic immigrants, there are differences by gender: female refugees are more likely to be married, live in a larger household, and have a higher number of children in the household than male refugees.

The data in the table reveals that refugees have lower levels of education and language ability compared to economic immigrants. Economic immigrants appear to have significantly higher levels of education: economic immigrants have one more year of education on average, and refugees have significantly higher high school dropout rates. Refugees are rated 0.7 to 0.8 points lower in German-speaking ability.

In terms of labor market outcomes, refugees report worse employment and wage than economic immigrants. Male refugees are less than half as likely to be employed as male economic immigrants. Female refugees are more than three times less likely to be employed. In addition, among employed refugees, a higher fraction reports part-time status employment compared to economic immigrants. In terms of wages, refugees report lower hourly wages.

[Table 3 about here]

Table 4 compares the characteristics of economic immigrants and refugees by years since migration into Germany. A longer length of stay positively correlates with educational attainment and language ability. In terms of long-run outcomes, refugees with more extended stay have similar education and German fluency levels to comparable economic immigrants. Figure 1 provides additional information on this issue. It displays data on the average educational attainment and the trajectory of language ability over time in Germany for both economic immigrants and refugees.

[Table 4 about here]

When we examine the first ten years since migration in the first and second panels, we see that refugees with a longer length of stay in Germany are more educated than newer refugees. At the same time, the educational level of economic immigrants remains constant. The result is that both male and female refugees with 15 years in Germany have about the same educational level as comparable economic immigrants. One interpretation is that refugees make more educational gains over time than economic immigrants. This conclusion would be in line with findings from Cortes (2004). However, an alternative explanation is also plausible: older refugee cohorts have different levels of education than the new refugee cohort.

[Figure 1 about here]

The first panel implies that the latter explanation plays a significant role: although males and females follow a similar pattern in their first 15 years in Germany, male refugee education is negatively associated with years since migration after 15 years. Improvements in educational attainment are unlikely to explain this particular pattern: change in the compositions of the refugee cohorts is a more plausible explanation of the higher educational attainment over time. The third and fourth panels show that language ability positively correlates with time spent in Germany for both economic immigrants and refugees. The new refugees have worse German-speaking ability than new economic immigrants. However, older refugee cohorts have similar fluency levels.

## 4. Empirical Strategy

Next, we present our empirical approach, which allows us to investigate the determinants of wage growth between refugees and economic immigrants. Using the individual longitudinal SOEP data, we conduct our empirical analysis on a sample of immigrants (refugees and economic immigrations) in Germany, ages 16-65, surveyed between 1984 and 2018. We examine the determinants of earnings growth through a series of Mincer-style equations of the following form:

$$Y_{ict} = \alpha_0 + \alpha_1 D^{Refugee} + \mathbf{X}_{ict}\gamma + \alpha_2 \text{Language}_{ict} + \alpha_3 \text{Education}_{i,t} + \alpha_4 \text{Stay}_{ict} + \delta_s + \delta_t + \delta_s \times \delta_t + \delta_o + \varepsilon_{ict} \quad (1)$$

For $Y_{i,t}$, we use two outcomes: logged hourly wage and employment status (set to 1 if the person reports being employed, and zero otherwise). We used the regressions with a binary outcome using a linear probability model. The specification is estimated at the state level, $c$. $D^{Refugee}$ is a dummy variable, indicating a refugee status immigrant. $\text{Language}_{i,t}$ is a proxy variable for fluency in the German language. Although the raw German-language skills measure uses a Likert scale variable, we standardize this measure.[13] $Education_{i,t}$ is a vector of educational attainment indicators: whether the person completed a high-school education and completed more than high-school education; the reference group for the educational variable is the group of people having less than high-school. $Stay_{it}$ is the length of stay in the host country; the variable comprises four groups: up to five years, six to ten years, eleven to fifteen years, and more than fifteen years (less than five years is the base group). We also include a vector of demographic and socioeconomic characteristics, $\mathbf{X}_{i,t}$, which accounts for age, gender (binary indicator), whether the person is currently married, and the number of children in the household. Finally, we include state fixed effects $\delta_s$, year effects $\delta_t$, interactions of state and year, and country of origin fixed effects $\delta_o$.

We focus on several variables of interest. The first coefficient of interest is $\alpha_1$. It indicates how refugees perform relative to economic immigrants. We also examine the set of coefficients, γ, associated with the vector of socioeconomic control variables, which will shed light on critical constraints for wage growth between refugees and economic immigrants. We estimate (1) separately for men, women, and the entire sample.

---

[13] The Likert scales for German skills range from excellent (5), good (4), sufficient (3), poor (2), and none (1).

To examine for any heterogeneous effects related to immigration status and the duration of stay, we interact the refugee status and the length of stay (capturing the four duration binary indicators noted above) variables:

$$Y_{i,t} = \alpha_0 + \alpha_1 D^{Refugee} + \mathbf{X}_{i,t}\gamma + \alpha_2 \text{Language}_{i,t} + \alpha_3 \text{Education}_{i,t} + \alpha_4 \text{Stay}_{i,t} + \alpha_5 D^{Refugee} \times \text{stay}_{i,t} + \delta_s + \delta_t + \delta_s \times \delta_t + \delta_o + \varepsilon_{it} \quad (2)$$

The coefficient, $\alpha_5$, associated with the interactions captures any heterogeneous treatment effects for refugees based on different lengths of stay.

## 5. Results

### A. Downgrading By Immigration Group Status

We start by examining the magnitude of downgrading for each group of immigrants. Table 5 reports the results for two outcomes: employment status and earnings.

For both outcomes, the results reveal that refugees downgrade substantially more than other immigrants. This pattern is robust both for the male and the female samples. However, we see strong evidence of stronger downgrading among male refugees than female refugees. Overall, refugees are almost 9 percent less likely to be employed than other immigrants.

[Table 5 about here]

Conditional on being employed, refugees earn 15.8 percent less in hourly wages than comparable economic immigrants. Female refugees report 7.6 percent lower wages than female economic immigrants (although the estimate is insignificant, it is close to statistical significance at the 10-percent level). The effect size associated with the refugee status is substantial among the male refugees, who report 20.9 percent lower wages (significant at the 1-percent level) than comparable economic immigrants. The results suggest that the wage disparity between refugees and economic immigrants is more pronounced in the sample of men than women.

Table 5 shows that higher human capital for both immigrant groups exerts a significant influence on wages. We see that post-high school education raises the earnings of immigrants by 36.6 percent. Female immigrants are 23 percent less likely to be employed. Employed female immigrants earn 28 percent lower wages. One somewhat unusual pattern is that male and female immigrants earn lower wages and are less likely to be employed the more children they have. However, the effect on employment is more substantial for women (-1.9 percent for men vs. -6.8 percent for women). A second result relates to how marital status affects earnings and particularly the effect of marital status on the earnings of female immigrants: married female immigrants are 7.6 percent less likely to be employed. However, those who are employed earn 3.9 percent more than their unmarried counterparts.

Language ability also plays a unique role in immigrant labor market outcomes. Across the overall immigrant sample (economic immigrants and refugees combined), one standard deviation (SD) increase in German-speaking ability is associated with a 5.8 percent higher chance of employment for men and a 7.6 percent higher chance for women. The same increase is also associated with a 4.4 percent wage increase for men and a 5.6 percent wage increase for women.

Finally, it appears that both employment probability and wages improve for the overall immigrant sample as time in Germany increases. Immigrants who have been in Germany for 6-10 years are 10.3 percent more likely to be employed than more recent arrivals with similar characteristics. They also earn 4.8 percent higher hourly wages.

B. Convergence Over Time

We next examine how refugee outcomes change over time relative to other immigrants. Figure 2 plots the change in employment and wages over time for refugees and economic immigrants. Without controlling for any socioeconomic variables, new refugees do significantly worse than new economic immigrants. However, refugees and economic immigrants who have been in Germany longer exhibit similar employment levels. Additionally, the third panel indicates that employed male refugees exhibit lower wages than their economic counterparts, no matter how long their stay in Germany. The same panel also implies that employed female refugees who have been in Germany for longer than 20 years have similar wages as female economic immigrants.

[Figure 2 about here]

Next, and based on additional specifications, we investigate further the trends detailed in Figure 2. Specifically, we examine how refugees and other immigrants differ in their returns to education and language skills. We do so by estimating three separate specifications—the first specification includes interactions between the refugee status and years spent within Germany. The second specification also includes controls for interactions between refugee status and educational attainment. The third specification includes the controls from the first two specifications and interaction terms between the refugee status and language ability. We report the results based on these specifications in Tables 6 and 7.

[Table 6 about here]

The results from Table 6 show that male refugees are 31.7 percent less likely to be employed than comparable male economic immigrants within five years of immigration. However, female refugees are only 16.1 percent less likely to be employed than their economic counterparts. In order to test whether this result is due to gender asymmetries between refugees and economic immigrants in education or language, we control for educational attainment and language ability conditional on refugee status (columns 4-9). The result does not change, indicating that this difference is not due to gender asymmetries between refugees and economic immigrants.

Table 6 also provides evidence that refugees converge towards economic immigrants in employment over time. The coefficient for refugee status in Column 1 indicates that refugees have 23.8 percent lower odds of employment in the first five years in Germany. However, the coefficients for refugee status interacted with years in Germany (columns 1-3) indicate that refugees close the entire gap after ten to fifteen years. We find a 4.4 percent gap for those in Germany for longer than fifteen years, but this gap is small and not statistically significant. The gap after fifteen years is small and insignificant for men. A seven percent gap remains for women without accounting for refugee education and refugee language ability.

Refugees with more than a high school education are 9 percent less likely to be employed than their economic contemporaries. Interestingly, this effect size is smaller and insignificant for men; this estimate is 11.3 percent and statistically significant for women. This discrepancy is likely because highly educated female refugees who cannot find a job commensurate with their abilities likely exit the labor force. In contrast, highly educated male refugees will find a less skilled occupation.

Male refugees also receive a modest premium in language fluency: one SD increase in German spoken proficiency raises their odds of employment by seven percent relative to male economic immigrants. Results from specifications (2) and (3) (columns 4-9) show that refugees with more than a high school education find it more difficult to become employed, possibly due to the barriers to high-skill employment for refugees.

[Table 7 about here]

Table 7 reports the regression results for log hourly wages. We include the same three specifications for controls as in Table 6. Conditional on being employed, female refugees in their first five years in Germany downgrade more than male refugees. Female refugees receive a 26.6 percent penalty in the first five years, whereas male refugees only receive a 17-percent penalty. The gap between female and male refugees persists after controlling for educational attainment and language ability (columns 4-9).

However, female refugees quickly assimilate in earnings to their refugee counterparts. After six to ten years in Germany, there is no longer a statistically significant gap between female refugees and female economic immigrants (column 3). Meanwhile, the earnings gap for male refugees fails to shrink, and it even increases from 17 percent to 30.9 percent eleven to fifteen years after migration into Germany.

The second specification adds interactions between refugee status and education (columns 3-6). We find that, on average, refugees with more than a high school education face an additional 25.7 percent wage penalty. This result coincides with the prediction that high-skilled refugees are less prepared for the new labor market and face more legal restrictions to high-skilled positions than their economic immigrant contemporaries. They receive lower returns to their accumulated human capital and likely take lower-skill jobs. Refugees with a high school education or less face no additional penalty, likely because similarly educated economic immigrants will also work in low-skill occupations. In the third specification (columns 7-9), we add an interaction between refugee status and language ability. The coefficient is 2.5 percent and is not statistically significant. We thus find no evidence that the wage returns to language ability are different for refugees and economic immigrants.

A noticeable pattern in Tables 6 and 7 is that women's initial economic immigrant-refugee gap is lower for employment status but higher for earnings. Furthermore, although both male and female refugees close the gap in employment over time, female refugees completely catch up to female economic immigrants in terms of earnings. However, among men, the gap between refugees and economic immigrants does not close. Different cohorts may muddle the results from table 7 over time.

However, if we restrict the sample to more recent cohorts and repeat the same specifications as just noted, we see the same pattern: only women close the refugee wage gap.[14]

C. Investigating Potential Mechanisms

Next, we investigate more deeply some of the results in Tables 6 and 7. We specifically focus on potential mediating channels underlying the outcomes we analyze in these two tables. Refugees will likely catch up more quickly if they improve their human capital relative to economic immigrants. Recall from Figure 1 that a longer length of stay in Germany correlates with higher language ability and education for refugees. These variables are likely two mediating channels for changes to the reported labor market outcomes. Language acquisition can increase with time spent in the host country, especially during the first few years. In contrast, improvements in educational attainment in late adulthood is an unlikely story. This subsection examines the change in German-speaking skills and education over time for refugees and economic immigrants. We estimate the following model:

$$M_{it} = \gamma_1 YSM_{it} + \gamma_2 YSM_{it}^2 + \gamma_3 Refugee + \gamma_4 Female + \gamma_5 Ref \times Fem + Ref \times YSM_{it} + \gamma_6 Ref \times Fem \times YSM_{it} + \delta_c \quad (3)$$

*M* is the variable of interest (language skills or educational attainment) likely mediating the channels on the outcomes in Tables 6 and 7. $YSM_{it}$ measures the years spent within Germany. We estimate (3) for two groups, those with five or fewer years in Germany and those with more than five years in Germany. In order to determine whether these apparent changes are actually due to differences in cohorts, we also run a fixed effect model via an unbalanced panel consisting of individuals with multiple observations over time.

---

[14] Ideally, we would restrict the analysis to separate cohorts. Unfortunately, we are limited by the data, as the only well-populated refugee cohort is the primarily Syrian group that entered Germany in 2013 or later. The economic immigrant cohort over this period is not suffiently populated (see Table 2) in order for us to find statistical significance, considering the large number of fixed effects and clustered standard errors. Additionally, since we only have outcomes for this group until 2018, we may not see a full picture of convergence over time. However, if we run a regression on this sample with a refugee ×YSM interaction, we find a coefficient of 0.136 for women with a p value of 0.192, implying possible convergence. Furthermore, if we limit the sample to those who immigrated into Germany in 2003 or later, we find statistically significant results similar to those in Table 7.

Table 8 reports the results based on specification (3). In the first five years after migration, each year in Germany increases speaking ability by 0.41 SDs. This improvement is smaller for observations after five years in Germany. This result is consistent with an interpretation that immigrants likely learn sufficient German within their first few years in the country. During the first five years, we see that male refugees start 0.14 SDs below economic immigrants in language ability but converge towards economic immigrants by 0.07 SDs per year. Female refugees, meanwhile, converge more slowly in speaking ability, at 0.48 SDs per year.

[Table 8 about here]

Each year in Germany is associated with a relatively small 0.02 increase in years of education. Male refugees who have lived in Germany for more than five years have 0.33 fewer years of education. Female refugees have the same educational levels as their economic counterparts. It also appears that refugees increase their education faster than economic immigrants, especially males in the first five years (0.133 more years of education per year than male economic immigrants). Next, we turn to the issue, examining whether these estimations reflect changes within cohorts or differences between cohorts.

Table 9 reports the panel level coefficients for years since migration, separated by gender and five years since migration. According to this table, refugees improve their German in their first five years at a much higher rate than economic immigrants. This result supports the notion from Figure 1 that refugees converge in language ability. However, this table also reveals that, contrary to the implications in Figure 1 and Table 8, refugees increase their education more slowly than economic immigrants in the first five years. However, refugees and economic immigrants who report being in Germany for longer than five years gain educational improvements at similar rates. Thus, we see that while the distribution in language by years since migration displayed in Figure 1 is primarily due to increased fluency over time, the distribution in education is due to differences across cohorts.

[Table 9 about here]

Tables 8 and 9 provide explanations for the convergence results. As noted earlier (based on Tables 6 and 7), fluency is more critical for refugees in finding a job than earning higher wages conditional on having a job. Since refugees converge in fluency, it seems reasonable that they would converge in employment but not necessarily in wages. Likewise, since refugees do not converge in education, they will not meet the qualification standards for high-skilled labor in the new market and must downgrade in occupation. This implies that education is not a mechanism in which refugees improve their wages, partially explaining why male refugees do not converge in wages over time. However, data limitations prevent us from exploring additional mechanisms to shed further light on why this pattern is different for female refugees.

## 6. Conclusion

In this study, we investigate the extent of skill downgrading among various immigrant groups in Germany. We specifically compare the magnitude of downgrading between refugees and economic immigrants. We also examine whether refugees catch up to economic immigrants on various labor market dimensions: employment status and wage outcomes and how the convergence trajectory varies by gender.

We find significant skill downgrading among both refugees and economic immigrants. However, we show that refugees downgrade more than economic immigrants for hourly wages. This downgrading differential between the two groups is considerably smaller for employment status. Second, we show that gender plays a critical role in how refugees converge to economic immigrants on various labor market outcomes. Male refugees do not converge to economic immigrants in terms of wages. In contrast, female refugees converge in wages reasonably quickly.

Our results indicate that language and educational attainment are critical determinants in immigrant labor market outcomes. Refugees initially exhibit lower levels of German-speaking ability and education than economic immigrants. However, we find that refugees improve their German language skills quickly. Our longitudinal analysis indicates that refugees do not improve their educational attainment faster relative to economic immigrants. Once we account for educational attainment, we find that new refugee immigrants are less likely to be employed than economic immigrants. Similarly, we show that refugees earn lower wages than new economic immigrants.

Over time, both male and female refugees catch up in employment status, although only female refugees converge in earnings towards their economic immigrant contemporaries. Convergence in employment but not wages may be explained by the increase in refugee language ability, as refugee fluency is crucial for refugee employment and not for refugee earnings conditional on employment. However, improvements in educational attainment cannot fully explain the wage convergence of female refugees.

# Figures and Tables

MALES · FEMALES

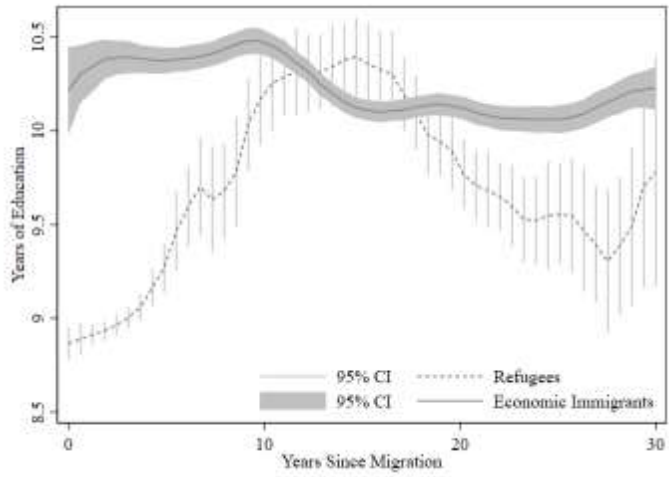
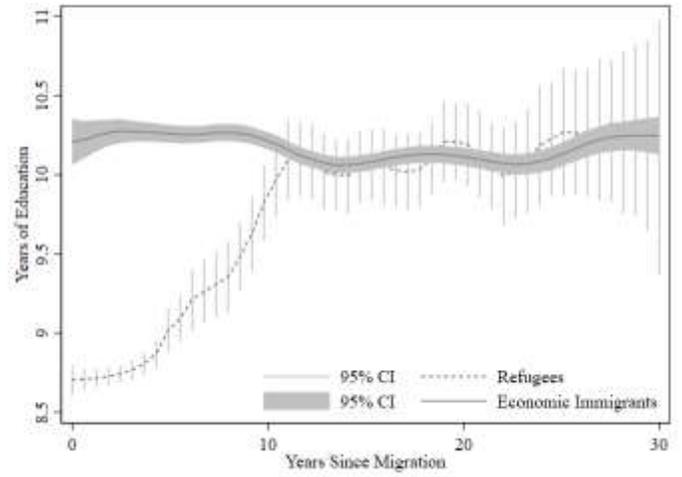
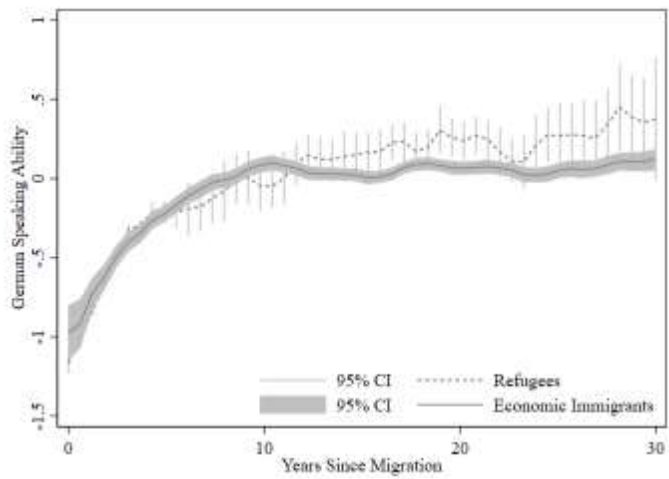
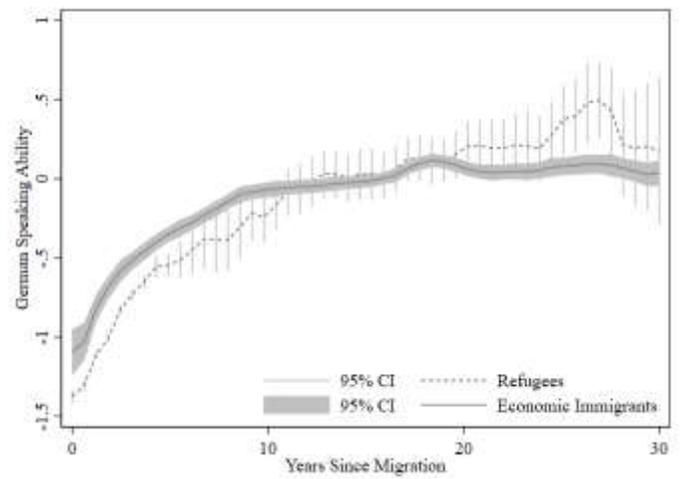

**Figure 1** – Educational attainment by years since migration in Germany. The top panels show years of education over one's lifetime by years in Germany. The bottom panels show standardized German-speaking ability. The left panels show outcomes by immigrant type; The right panels show outcomes by year of immigration into Germany.

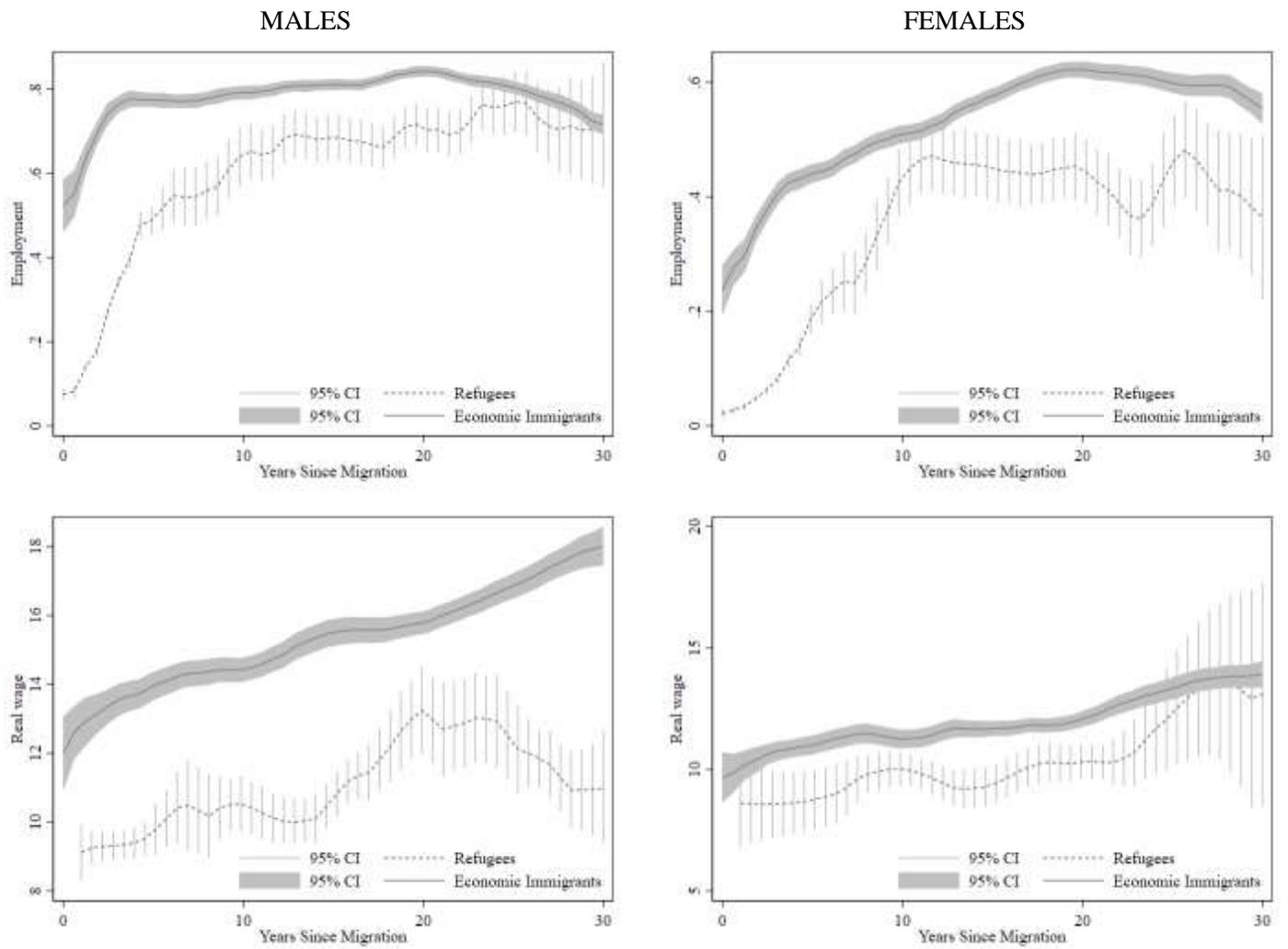

**Figure 2** – Employment and wages by years since migration in Germany. *Note:* The top panels show employment by years in Germany; the bottom panels show the real hourly wage. The left panels show outcomes by immigrant type; the right panel shows outcomes by immigration into Germany.

Table 1 – Immigrants by type and years in Germany

|  | Refugee | Economic Immigrants | Total |
|---|---|---|---|
| Males |  |  |  |
|   0-5 years | 8,784 | 3,223 | 12,007 |
|   6-10 years | 510 | 5,259 | 5,769 |
|   11-15 years | 660 | 7,122 | 7,782 |
|   16+ years | 1,725 | 28,033 | 29,758 |
|   Total | 11,679 | 43,637 | 55,316 |
| Females |  |  |  |
|   0-5 years | 5,819 | 4,301 | 10,120 |
|   6-10 years | 499 | 6,746 | 7,245 |
|   11-15 years | 534 | 8,317 | 8,851 |
|   16+ years | 1,306 | 27,939 | 29,245 |
|   Total | 8,158 | 47,303 | 55,461 |

*Notes:* The table represents the number of sample observations in each category, including multiple observations from the same immigrants for different years. Data is unweighted. The data covers the period 1984-2018. Economic immigrants include immigration for specific reasons (family, new job, etc). Observations are ages 16-65.

**Table 2** – Number of Immigrants by country of origin and migration period

|  | Before 1995 | 1995-2002 | 2003-2010 | 2011-2018 | Total |
|---|---|---|---|---|---|
| Refugees | | | | | |
| Syria | 10 | 31 | 17 | 4,081 | 4,139 |
| Iraq | 2 | 65 | 72 | 1,122 | 1,261 |
| Afghanistan | 23 | 16 | 16 | 997 | 1,052 |
| Other | 509 | 405 | 131 | 1,807 | 2,852 |
| Total | 544 | 517 | 236 | 8,007 | 9,304 |
| Economic | | | | | |
| Turkey | 1,833 | 259 | 122 | 24 | 2,238 |
| Greece | 522 | 56 | 41 | 59 | 678 |
| Italy | 773 | 106 | 46 | 59 | 984 |
| Spain | 415 | 32 | 40 | 44 | 531 |
| Romania | 264 | 110 | 216 | 194 | 784 |
| Poland | 740 | 224 | 293 | 201 | 1,458 |
| Russia | 429 | 613 | 225 | 35 | 1,302 |
| Kazakhstan | 436 | 597 | 142 | 21 | 1,196 |
| Other | 1,995 | 913 | 844 | 683 | 4,435 |
| Total | 7,407 | 2,910 | 1,969 | 1,320 | 13,606 |

*Notes:* Economic immigrants include immigration for specific reasons (e.g., family, new job, etc.). The statistics are unweighted. All observations represent a unique immigrant ages 16-65 when surveyed. Individuals are organized based on the year of migration to Germany.

**Table 3** – Descriptive Statistics (Means)

|  | Native | | Refugee | | Economic | |
|---|---|---|---|---|---|---|
|  | Males | Females | Males | Females | Males | Females |
| **Dependent Variables** | | | | | | |
| Employed | | | | | | |
|   All | 0.78 | 0.64 | 0.35 | 0.16 | 0.78 | 0.55 |
|   Full-time | 0.68 | 0.31 | 0.21 | 0.05 | 0.71 | 0.28 |
|   Part-time | 0.05 | 0.30 | 0.10 | 0.10 | 0.04 | 0.25 |
|   Vocational Training | 0.04 | 0.03 | 0.04 | 0.02 | 0.03 | 0.02 |
| Log Hourly Wage | | | | | | |
|   All | 2.69 | 2.45 | 2.20 | 2.13 | 2.63 | 2.36 |
|   Full-time | 2.80 | 2.58 | 2.39 | 2.40 | 2.70 | 2.48 |
|   Part-time | 2.34 | 2.45 | 1.94 | 2.09 | 2.21 | 2.30 |
|   Vocational Training | 1.33 | 1.33 | 1.44 | 1.45 | 1.40 | 1.33 |
| **Control Variables** | | | | | | |
| Age | 40.66 | 40.74 | 34.64 | 34.81 | 41.29 | 40.53 |
| Married | 0.57 | 0.57 | 0.55 | 0.65 | 0.72 | 0.72 |
| Education | | | | | | |
|   Years of Education | 12.40 | 12.19 | 9.23 | 9.10 | 10.37 | 10.25 |
|   High School Dropout | 0.10 | 0.15 | 0.65 | 0.68 | 0.37 | 0.45 |
|   High School Only | 0.64 | 0.63 | 0.15 | 0.14 | 0.49 | 0.38 |
|   Post-HS Education | 0.26 | 0.22 | 0.20 | 0.18 | 0.14 | 0.18 |
| Years since Migration | | | | | | |
|   Total Years | | | 5.90 | 6.35 | 19.54 | 18.22 |
|   0-5 Years | | | 0.75 | 0.71 | 0.07 | 0.09 |
|   6-10 Years | | | 0.04 | 0.06 | 0.12 | 0.14 |
|   11-15 Years | | | 0.06 | 0.07 | 0.16 | 0.18 |
|   16+ Years | | | 0.15 | 0.16 | 0.64 | 0.59 |
| German Speaking Ability | 4.73 | 4.78 | 3.14 | 2.83 | 3.74 | 3.64 |
| Number of Children | 0.73 | 0.78 | 1.66 | 2.14 | 1.07 | 1.09 |
| Household Members | 3.06 | 3.00 | 3.87 | 4.61 | 3.55 | 3.49 |

*Notes:* Observations are ages 16-65, unweighted. The hourly wage is in Euro, with a base year of 2015. German-speaking ability is standardized from a discrete scale of 1-5. Years since migration, refugee and marital status, and education groups are binary variables. Number of children denotes children within the household only

| | 0-5 | | 6-10 | | 11-15 | | 16+ | |
|---|---|---|---|---|---|---|---|---|
| | Refugee | Economic | Refugee | Economic | Refugee | Economic | Refugee | Economic |
| Dependent Variables | | | | | | | | |
| Employed | | | | | | | | |
|    All | 0.18 | 0.54 | 0.45 | 0.61 | 0.58 | 0.67 | 0.59 | 0.69 |
|    Full-time | 0.07 | 0.38 | 0.25 | 0.42 | 0.31 | 0.47 | 0.39 | 0.53 |
|    Part-time | 0.07 | 0.14 | 0.16 | 0.16 | 0.23 | 0.16 | 0.19 | 0.15 |
|    Vocational Training | 0.04 | 0.03 | 0.04 | 0.04 | 0.04 | 0.04 | 0.01 | 0.01 |
| Log Hourly Wage | | | | | | | | |
|    All | 2.03 | 2.35 | 2.17 | 2.39 | 2.13 | 2.43 | 2.35 | 2.59 |
|    Full-time | 2.26 | 2.49 | 2.39 | 2.56 | 2.37 | 2.58 | 2.48 | 2.68 |
|    Part-time | 1.92 | 2.15 | 1.98 | 2.20 | 1.96 | 2.25 | 2.14 | 2.34 |
|    Vocational Training | 1.49 | 1.38 | 1.48 | 1.36 | 1.27 | 1.32 | 1.33 | 1.44 |
| Control Variables | | | | | | | | |
| Age | 32.87 | 32.88 | 35.74 | 34.82 | 38.06 | 36.52 | 42.45 | 44.47 |
| Married | 0.55 | 0.67 | 0.69 | 0.69 | 0.67 | 0.68 | 0.75 | 0.76 |
| Education | | | | | | | | |
|    Years of Education | 8.89 | 10.31 | 9.65 | 10.34 | 10.22 | 10.15 | 9.99 | 10.27 |
|    High School | 0.72 | 0.42 | 0.60 | 0.40 | 0.44 | 0.43 | 0.51 | 0.41 |
|    High School Only | 0.07 | 0.30 | 0.22 | 0.38 | 0.37 | 0.41 | 0.36 | 0.46 |
|    Post-HS Education | 0.21 | 0.28 | 0.18 | 0.22 | 0.19 | 0.16 | 0.13 | 0.12 |
| German Speaking Ability | 2.87 | 3.15 | 3.44 | 3.58 | 3.75 | 3.68 | 3.93 | 3.80 |
| Number of Children | 1.90 | 1.00 | 1.96 | 1.26 | 1.61 | 1.37 | 1.68 | 1.01 |
| Household Members | 4.06 | 3.34 | 4.45 | 3.64 | 4.42 | 3.82 | 4.52 | 3.48 |

Table 4 – Descriptive Statistics By Years in Germany

*Notes:* Observations ages 16-65, unweighted. Hourly wage is in Euro, base year 2015. German-speaking ability is standardized from a discrete scale of 1-5. Years since migration, refugee and marital status, and education groups are binary variables. The number of children is within the household only.

**Table 5** – Employment Status and Hourly Wages

|  | Employment (=1 if yes) | | | Hourly Wages (Logged) | | |
|---|---|---|---|---|---|---|
|  | All | Males | Females | All | Males | Females |
|  | (1) | (2) | (3) | (4) | (5) | (6) |
| Refugee | -0.085*** | -0.100*** | -0.069*** | -0.158*** | -0.209*** | -0.076 |
|  | (0.017) | (0.025) | (0.022) | (0.018) | (0.016) | (0.049) |
| 6-10 Years | 0.103*** | 0.112*** | 0.103*** | 0.048*** | 0.038* | 0.064** |
|  | (0.011) | (0.016) | (0.008) | (0.013) | (0.021) | (0.029) |
| 11-15 Years | 0.145*** | 0.126*** | 0.165*** | 0.092*** | 0.100*** | 0.089*** |
|  | (0.013) | (0.019) | (0.011) | (0.012) | (0.016) | (0.022) |
| >15 Years | 0.149*** | 0.117*** | 0.178*** | 0.146*** | 0.148*** | 0.147*** |
|  | (0.021) | (0.022) | (0.023) | (0.018) | (0.019) | (0.030) |
| Age | 0.048*** | 0.045*** | 0.046*** | 0.068*** | 0.069*** | 0.065*** |
|  | (0.001) | (0.002) | (0.002) | (0.003) | (0.002) | (0.005) |
| Age Squared | -0.001*** | -0.001*** | -0.001*** | -0.001*** | -0.001*** | -0.001*** |
|  | (0.000) | (0.000) | (0.000) | (0.000) | (0.000) | (0.000) |
| Female | -0.226*** |  |  | -0.277*** |  |  |
|  | (0.008) |  |  | (0.009) |  |  |
| High School Graduate | 0.025*** | 0.024*** | 0.023* | 0.105*** | 0.086*** | 0.123*** |
|  | (0.007) | (0.007) | (0.013) | (0.008) | (0.009) | (0.016) |
| Post-HS Education | 0.052*** | 0.046*** | 0.055** | 0.366*** | 0.376*** | 0.346*** |
|  | (0.012) | (0.009) | (0.021) | (0.021) | (0.027) | (0.027) |
| German Speaking Ability | 0.072*** | 0.058*** | 0.076*** | 0.053*** | 0.044*** | 0.056*** |
|  | (0.004) | (0.004) | (0.006) | (0.007) | (0.008) | (0.008) |
| Married | -0.020** | 0.042*** | -0.078*** | 0.095*** | 0.144*** | 0.039** |
|  | (0.007) | (0.011) | (0.009) | (0.011) | (0.013) | (0.016) |
| Number of Children | -0.038*** | -0.019*** | -0.068*** | -0.019*** | -0.020*** | -0.027** |
|  | (0.003) | (0.004) | (0.003) | (0.003) | (0.004) | (0.011) |
| Baseline Mean | 0.588 | 0.686 | 0.489 | 2.491 | 2.596 | 2.347 |
| R-sq | 0.314 | 0.352 | 0.281 | 0.293 | 0.321 | 0.227 |
| N | 51571 | 27200 | 24371 | 27215 | 16408 | 10807 |

*Notes:* Standard errors are in parentheses. All regressions are immigrant sample only and control for state, survey year, state*survey year, and country of origin fixed effects. Employment is binary (0 = not employed, 1 = employed). Hourly wage is in Euro with base year 2015. Regressions for log hourly wage are conditional on employment. German-speaking ability is standardized from a discrete scale of 1-5 (5=excellent). Years since migration, refugee and marital status, and education groups are binary. The number of children only includes children within the household. All results are clustered by state.
*** Significant at the 1 percent level. ** Significant at the 5 percent level. * Significant at the 10 percent level.

**Table 6** – Employment

|  | All (1) | Males (2) | Females (3) | All (4) | Males (5) | Females (6) | All (7) | Males (8) | Females (9) |
|---|---|---|---|---|---|---|---|---|---|
| Refugee | -0.238*** | -0.317*** | -0.161*** | -0.212*** | -0.307*** | -0.125*** | -0.196*** | -0.268*** | -0.130*** |
|  | (0.028) | (0.028) | (0.033) | (0.030) | (0.026) | (0.038) | (0.029) | (0.028) | (0.036) |
| Refugee × 6-10 Years | 0.210*** | 0.248*** | 0.163*** | 0.208*** | 0.245*** | 0.161** | 0.197*** | 0.209*** | 0.163*** |
|  | (0.030) | (0.023) | (0.055) | (0.030) | (0.022) | (0.056) | (0.031) | (0.024) | (0.055) |
| Refugee × 11-15 Years | 0.252*** | 0.314*** | 0.192*** | 0.251*** | 0.309*** | 0.194*** | 0.233*** | 0.256*** | 0.198*** |
|  | (0.031) | (0.030) | (0.044) | (0.032) | (0.034) | (0.044) | (0.032) | (0.034) | (0.045) |
| Refugee × >15 Years | 0.194*** | 0.291*** | 0.091*** | 0.189*** | 0.284*** | 0.088*** | 0.166*** | 0.221*** | 0.094*** |
|  | (0.036) | (0.042) | (0.027) | (0.038) | (0.046) | (0.024) | (0.037) | (0.048) | (0.024) |
| 6-10 Years | 0.053*** | 0.030** | 0.077*** | 0.054*** | 0.030** | 0.078*** | 0.057*** | 0.041*** | 0.077*** |
|  | (0.012) | (0.014) | (0.011) | (0.012) | (0.014) | (0.011) | (0.012) | (0.014) | (0.011) |
| 11-15 Years | 0.092*** | 0.039* | 0.137*** | 0.094*** | 0.041* | 0.139*** | 0.099*** | 0.055** | 0.138*** |
|  | (0.016) | (0.019) | (0.014) | (0.016) | (0.020) | (0.014) | (0.016) | (0.021) | (0.015) |
| >15 Years | 0.101*** | 0.034 | 0.157*** | 0.104*** | 0.036 | 0.159*** | 0.110*** | 0.054* | 0.158*** |
|  | -0.238*** | -0.317*** | -0.161*** | -0.212*** | -0.307*** | -0.125*** | -0.196*** | -0.268*** | -0.130*** |
| Female | -0.226*** |  |  | -0.226*** |  |  | -0.225*** |  |  |
|  | (0.008) |  |  | (0.008) |  |  | (0.008) |  |  |
| High School Graduate | 0.023*** | 0.022*** | 0.022* | 0.031*** | 0.024*** | 0.033** | 0.034*** | 0.030*** | 0.032** |
|  | (0.007) | (0.006) | (0.013) | (0.009) | (0.007) | (0.015) | (0.009) | (0.007) | (0.015) |
| Post-HS Education | 0.049*** | 0.041*** | 0.053** | 0.083*** | 0.062*** | 0.086** | 0.087*** | 0.074*** | 0.085** |
|  | (0.012) | (0.009) | (0.021) | (0.021) | (0.014) | (0.029) | (0.021) | (0.015) | (0.029) |
| Refugee × HS |  |  |  | -0.018 | 0.007 | -0.036 | -0.025** | -0.006 | -0.033 |
|  |  |  |  | (0.010) | (0.024) | (0.033) | (0.011) | (0.023) | (0.033) |
| Refugee × Post-HS |  |  |  | -0.090*** | -0.044 | -0.113*** | -0.101*** | -0.076** | -0.110*** |
|  |  |  |  | (0.027) | (0.025) | (0.032) | (0.028) | (0.026) | (0.033) |
| German Speaking Ability | 0.072*** | 0.058*** | 0.075*** | 0.072*** | 0.058*** | 0.075*** | 0.064*** | 0.033*** | 0.076*** |
|  | (0.004) | (0.003) | (0.006) | (0.004) | (0.003) | (0.006) | (0.005) | (0.004) | (0.008) |
| Refugee × Speaking Ability |  |  |  |  |  |  | 0.027*** | 0.080*** | -0.007 |
|  |  |  |  |  |  |  | (0.006) | (0.005) | (0.008) |
| Baseline Mean | 0.588 | 0.686 | 0.489 | 0.588 | 0.686 | 0.489 | 0.588 | 0.686 | 0.489 |
| R-sq | 0.317 | 0.357 | 0.283 | 0.318 | 0.357 | 0.284 | 0.318 | 0.361 | 0.284 |
| N | 51,571 | 27,200 | 24,371 | 51,571 | 27,200 | 24,371 | 51,571 | 27,200 | 24,371 |

*Notes:* Standard errors are in parentheses. All regressions are immigrant sample only and control for state, survey year, state*survey year, and country of origin fixed effects. Employment is binary (0 = not employed, 1 = employed). German-speaking ability is standardized from a discrete scale of 1-5 (5=excellent). Years since migration, refugee and marital status, and education groups are binary. Number of children only includes children within the household. All results are clustered by state.
*** Significant at the 1 percent level. ** Significant at the 5 percent level. * Significant at the 10 percent level.

**Table 7** – Hourly Wages (Logged)

|  | All | Males | Females | All | Males | Females | All | Males | Females |
|---|---|---|---|---|---|---|---|---|---|
|  | (1) | (2) | (3) | (4) | (5) | (6) | (7) | (8) | (9) |
| Refugee | -0.193*** | -0.170*** | -0.266*** | -0.124*** | -0.098* | -0.195*** | -0.121** | -0.095* | -0.189*** |
|  | (0.041) | (0.046) | (0.061) | (0.040) | (0.047) | (0.054) | (0.042) | (0.050) | (0.053) |
| Refugee × 6-10 Years | 0.032 | -0.068 | 0.240** | 0.028 | -0.073 | 0.235*** | 0.027 | -0.078 | 0.239*** |
|  | (0.051) | (0.041) | (0.091) | (0.052) | (0.044) | (0.079) | (0.052) | (0.046) | (0.077) |
| Refugee × 11-15 Years | -0.034 | -0.139** | 0.170** | -0.036 | -0.137** | 0.155** | -0.043 | -0.147** | 0.150** |
|  | (0.035) | (0.059) | (0.069) | (0.042) | (0.059) | (0.064) | (0.042) | (0.062) | (0.063) |
| Refugee × >15 Years | 0.074 | -0.007 | 0.231*** | 0.056 | -0.019 | 0.188*** | 0.043 | -0.032 | 0.171*** |
|  | (0.058) | (0.071) | (0.053) | (0.062) | (0.072) | (0.046) | (0.069) | (0.083) | (0.048) |
| 6-10 Years | 0.043** | 0.047* | 0.046* | 0.044** | 0.048** | 0.047* | 0.044** | 0.049** | 0.047* |
|  | (0.016) | (0.022) | (0.026) | (0.016) | (0.022) | (0.026) | (0.016) | (0.022) | (0.026) |
| 11-15 Years | 0.092*** | 0.114*** | 0.075*** | 0.093*** | 0.118*** | 0.076*** | 0.094*** | 0.120*** | 0.076*** |
|  | (0.014) | (0.019) | (0.020) | (0.013) | (0.019) | (0.019) | (0.014) | (0.020) | (0.020) |
| >15 Years | 0.138*** | 0.152*** | 0.130*** | 0.141*** | 0.157*** | 0.132*** | 0.142*** | 0.159*** | 0.133*** |
|  | (0.021) | (0.026) | (0.027) | (0.022) | (0.026) | (0.027) | (0.022) | (0.027) | (0.027) |
| Female | -0.277*** |  |  | -0.277*** |  |  | -0.277*** |  |  |
|  | (0.009) |  |  | (0.008) |  |  | (0.008) |  |  |
| High School Graduate | 0.105*** | 0.087*** | 0.122*** | 0.113*** | 0.099*** | 0.123*** | 0.113*** | 0.099*** | 0.123*** |
|  | (0.008) | (0.009) | (0.016) | (0.011) | (0.010) | (0.019) | (0.012) | (0.010) | (0.019) |
| Post-HS Education | 0.366*** | 0.377*** | 0.345*** | 0.402*** | 0.436*** | 0.360*** | 0.403*** | 0.437*** | 0.361*** |
|  | (0.021) | (0.027) | (0.027) | (0.023) | (0.029) | (0.027) | (0.023) | (0.029) | (0.027) |
| Refugee × HS |  |  |  | -0.028 | -0.044 | 0.030 | -0.033 | -0.045 | 0.013 |
|  |  |  |  | (0.042) | (0.038) | (0.073) | (0.042) | (0.037) | (0.073) |
| Refugee × Post-HS |  |  |  | -0.257*** | -0.282*** | -0.256*** | -0.265*** | -0.288*** | -0.275*** |
|  |  |  |  | (0.038) | (0.038) | (0.068) | (0.038) | (0.037) | (0.076) |
| German Speaking Ability | 0.053*** | 0.044*** | 0.056*** | 0.053*** | 0.043*** | 0.057*** | 0.050*** | 0.040*** | 0.055*** |
|  | (0.007) | (0.008) | (0.008) | (0.006) | (0.007) | (0.008) | (0.006) | (0.006) | (0.008) |
| Refugee × Speaking Ability |  |  |  |  |  |  | 0.025 | 0.027 | 0.031 |
|  |  |  |  |  |  |  | (0.022) | (0.031) | (0.023) |
| Baseline Mean | 2.491 | 2.596 | 2.347 | 2.491 | 2.596 | 2.347 | 2.491 | 2.596 | 2.347 |
| R-sq | 0.293 | 0.322 | 0.227 | 0.295 | 0.325 | 0.229 | 0.296 | 0.325 | 0.229 |
| N | 27,215 | 16,408 | 10,807 | 27,215 | 16,408 | 10,807 | 27,215 | 16,408 | 10,807 |

*Notes:* Standard errors are in parentheses. All regressions are immigrant sample only (conditional on employment) and control for state, survey year, state*survey year, and country of origin fixed effects. Hourly wage is in Euro with base year 2015. German-speaking ability is standardized from a discrete scale of 1-5 (5=excellent). Years since migration, refugee and marital status, and education groups are binary. The number of children only includes children within the household. All results are clustered by state.

\*\*\* Significant at the 1 percent level. \*\* Significant at the 5 percent level. \* Significant at the 10 percent level.

## Table 8 – Impact on Mechanism Variables

|  | GERMAN-SPEAKING ABILITY | | | YEARS OF EDUCATION | | |
|---|---|---|---|---|---|---|
|  | All | 5 or Less | > 5 Years | All | 5 or Less | > 5 Years |
|  | (1) | (2) | (3) | (4) | (5) | (6) |
| YSM | 0.052*** | 0.413*** | 0.026*** | 0.017** | -0.072 | 0.021** |
|  | (0.003) | (0.030) | (0.004) | (0.008) | (0.048) | (0.010) |
| YSM Squared | -0.001*** | -0.043*** | -0.000*** | 0.000 | 0.009 | 0.000 |
|  | (0.000) | (0.004) | (0.000) | (0.000) | (0.008) | (0.000) |
| Refugee | -0.113*** | -0.136*** | -0.113*** | -0.303*** | -0.108 | -0.326*** |
|  | (0.018) | (0.041) | (0.017) | (0.045) | (0.075) | (0.042) |
| Female | -0.359*** | -0.256** | -0.184** | -0.893*** | -1.015*** | 0.056 |
|  | (0.060) | (0.099) | (0.063) | (0.152) | (0.173) | (0.304) |
| Refugee × Female | -0.258*** | -0.147** | -0.166** | -0.016 | -0.093 | -0.342 |
|  | (0.032) | (0.055) | (0.063) | (0.052) | (0.070) | (0.263) |
| Refugee × YSM | 0.020*** | 0.073*** | 0.009*** | 0.037*** | 0.133*** | -0.012 |
|  | (0.002) | (0.023) | (0.003) | (0.009) | (0.025) | (0.015) |
| Ref ×Female × YSM | 0.013*** | -0.025* | 0.010* | 0.013* | -0.029 | 0.029** |
|  | (0.003) | (0.013) | (0.005) | (0.006) | (0.021) | (0.014) |
| R-sq | 0.237 | 0.187 | 0.158 | 0.239 | 0.215 | 0.238 |
| N | 60,324 | 17,855 | 42,469 | 104,722 | 20,076 | 84,646 |

*Notes:* Standard errors are in parentheses. YSM=years since migration. All regressions are immigrant sample only and control for state and country of origin fixed effects. German-speaking ability is standardized from a discrete scale of 1-5. YSM denotes the number of years living in Germany. All results are clustered by state.
\*\*\* Significant at the 1 percent level. \*\* Significant at the 5 percent level. \* Significant at the 10 percent level.

| | Table 9 – Impacts on German-Speaking Ability and Educational Attainment (Panel) | | | | | |
|---|---|---|---|---|---|---|
| | German-speaking Ability | | | Years of Education | | |
| | All | 5 or Less | > 5 Years | All | 5 or Less | > 5 Years |
| | (1) | (2) | (3) | (4) | (5) | (6) |
| Refugee | | | | | | |
|     All | 0.249*** | 0.627*** | 0.030** | 0.025*** | 0.004 | 0.051*** |
|     Male | 0.276*** | 0.651*** | 0.020 | 0.024*** | 0.009 | 0.047*** |
|     Female | 0.217*** | 0.590*** | 0.041** | 0.030* | -0.004 | 0.059* |
| Economic | | | | | | |
|     All | 0.039*** | 0.217*** | 0.026*** | 0.054*** | 0.066*** | 0.052*** |
|     Male | 0.030*** | 0.294*** | 0.020*** | 0.053*** | 0.036 | 0.051*** |
|     Female | 0.046*** | 0.171*** | 0.032*** | 0.054*** | 0.080** | 0.053*** |

*Notes:* Standard errors are in parentheses. All regressions are immigrant sample only and based on an unbalanced panel with fixed effects. German-speaking ability is standardized from a discrete scale of 1-5. All results are robust.
\*\*\* Significant at the 1 percent level. \*\* Significant at the 5 percent level. \* Significant at the 10 percent level.